\documentclass[pra,groupedaddress,a4paper,twocolumn , nofootinbib]{revtex4}
\usepackage{dcolumn}

\usepackage{amsmath,amsfonts,bbm, graphicx,color, enumerate,soul}
\def\>{\rangle}
\def\<{\langle}

\def\H{ {\cal H} }
\def\M{ {\cal M} }
\def\II{ {\cal I} }
\def\N{ {\cal N} }

\def\Tr{ \mbox{tr} }

\def\diag{ \mathrm{diag}}

\usepackage[normalem]{ulem}  
\renewcommand\sout{\bgroup \color{red} \ULdepth=-.5ex \ULset}

\newcommand{\bra}[1]{\langle {#1} |}
\newcommand{\ket}[1]{| {#1} \rangle}

\newcommand{\wt}[1]{\widetilde{#1}}

\begin{document}

\title{Exchange Fluctuation Theorem for correlated quantum systems} %

\author{Sania Jevtic$^1$}


\author{Terry Rudolph$^2$}
\author{David Jennings$^2$}

\affiliation{$^1$Mathematical Sciences, Room 501, John Crank Building, Brunel University, Uxbridge UB8 3PH, United Kingdom \\ $^2$Controlled Quantum Dynamics Theory Group, Level 12, EEE, Imperial College London, London SW7 2AZ, United Kingdom}



\author{Yuji Hirono}
%
\author{Shojun Nakayama}

\author{Mio Murao}
\affiliation{Department of Physics, The University of Tokyo, Hongo 7-3-1 Bunkyo-ku Tokyo 113-0033, Japan}

\begin{abstract}
We extend the Exchange Fluctuation Theorem for energy exchange between
thermal quantum systems beyond the assumption of molecular chaos, and
describe the non-equilibrium exchange dynamics of correlated quantum
states. The relation quantifies how the tendency for systems to
equilibrate is modified in high-correlation environments. Our results
elucidate the role of measurement disturbance for such scenarios. We
show a simple application by finding a semi-classical maximum work
theorem in the presence of correlations.
\end{abstract}
\date{\today}

\maketitle
\section{Introduction}

Fluctuation theorems describe non-equilibrium transformations of a
thermodynamic system and constitute a refinement of the second law of
thermodynamics, the most well-known incarnations being the work
fluctuation theorems due to Jarzynski and Crooks \cite{Jarzynski1,
Jarzynski2,Crooks}. However, in addition to focussing on the extraction
of mechanical work from a single system, an equally fundamental
topic is the thermodynamic tendency of multipartite systems to
equilibrate. The canonical example of this is heat exchange
between two thermal systems at different temperatures and leads us
instead to fluctuation theorems for heat that provide a
quantitative description of the fluctuations in energy exchange between
two hot bodies.

\begin{figure}[t!]
\includegraphics[trim=2cm 1cm 13cm 0cm, width=.28\textwidth]{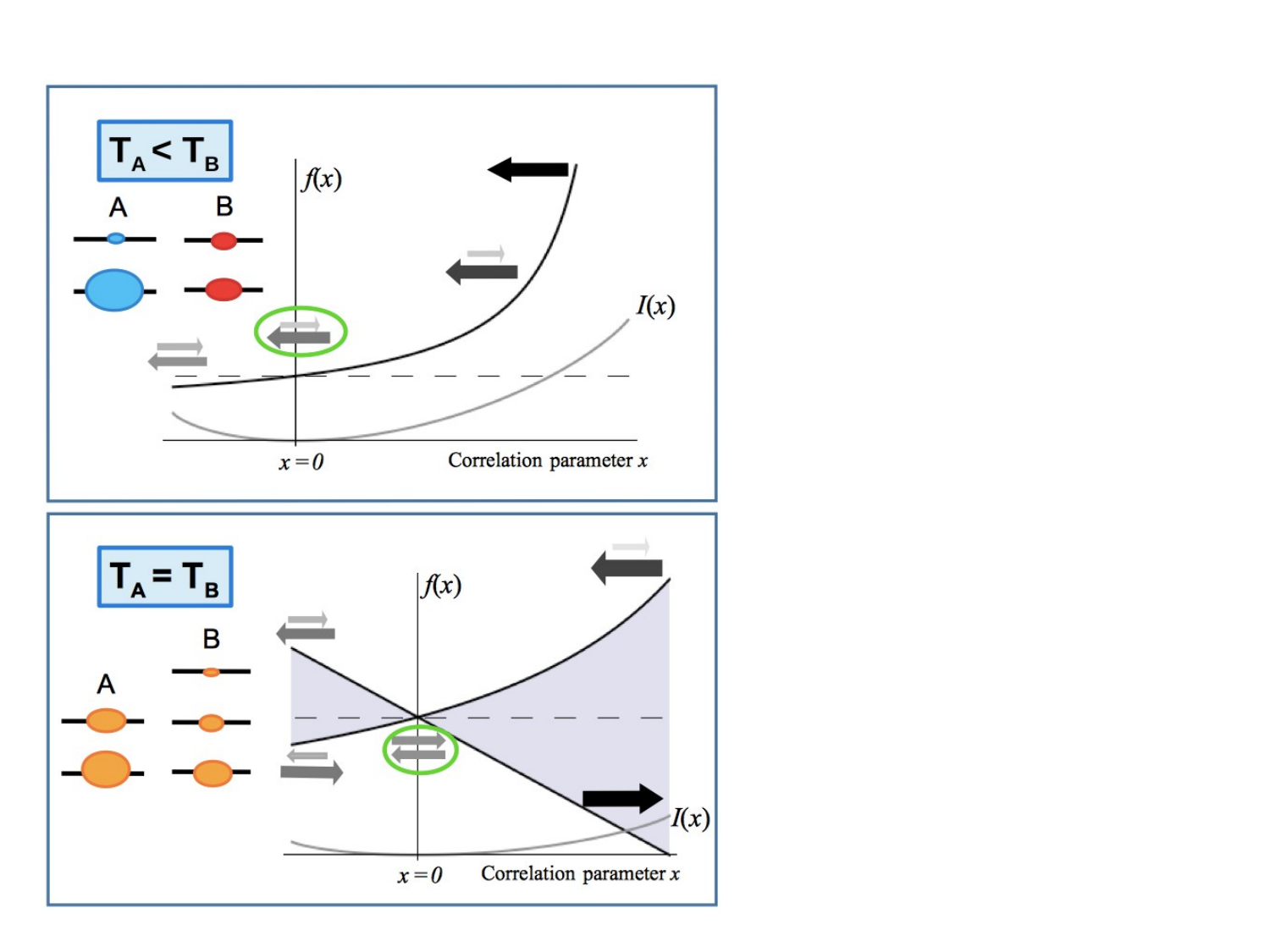} \centering
\caption{(Color online). \textbf{Why correlations matter:} Two systems $A$ and $B$ are at temperatures $T_A$ and $T_B$ initially. The generalised XFT, equation \eqref{genXFT}, is plotted as a function $f(x)$ of a correlation parameter $x$ and the mutual information $I(x)$ (light grey curve) is also included. As $I(x)$ increases, the XFT deviates from its value in the ``molecular chaos'' assumption where $x=0$ (dashed line). The likelihood of heat flow direction is indicated by the grey arrows for different $x$. A circle round the arrows indicates the uncorrelated regime $x=0$. For some $x$, backward heat flow can be completely suppressed. \textit{Top figure:} A two-qubit system where $B$ is hotter than $A$, $f(x)$ is a single black curve. \textit{Bottom figure:} A qubit-qutrit system at equal temperature. Unlike two-qubits, the XFT is only defined by a range (light grey region) and we see, even though $T_A = T_B$, correlations can induce heat flow on average, thereby violating the principle of detailed balance. This example is discussed in section V.
}
\label{Fig1}
\end{figure}

The thermodynamic arrow \cite{Huw-price-book, Zeh-book} is one
particular manifestation of the second law of thermodynamics, and in its
canonical form states that, on average, heat will flow from a hotter body
to a colder one. Specifically, given two thermal states
$\rho_A$ and $\rho_B$ at temperatures $T_A$ and $T_B$ with respect to
Hamiltonians $H_A$ and $H_B$, and an energy-conserving unitary evolution of the
joint state, $\rho_{A}\otimes \rho_B \rightarrow U \rho_A \otimes \rho_B
U^\dagger$, we define heat flow into $A$ as $Q_A=\Tr[H_A \rho'_A]
-\Tr[H_A\rho_A]$ where $\rho_A'$ is the final reduced state for $A$, and
we assume that the free Hamiltonians for $A$ and $B$ do not change. The
fact that Gibbsian states minimize the free energy yields the Clausius inequality
\begin{eqnarray}
\label{clausius}
Q_A \left(\frac{1}{T_A} - \frac{1}{T_B} \right) \ge 0,
\end{eqnarray}
and so if $T_A < T_B$ we have that the $Q_A$ is strictly positive, and
\emph{on average} energy is transferred from the hotter body to the
colder one. This is the standard thermodynamic arrow for heat flow.

However, a sharper expression of the directionality for heat flow exists
in the recent Exchange Fluctuation Theorem (XFT) due to Jarzynski and
W\'{o}jcik \cite{JarzynskiXFT}, which states that for two systems $A$
and $B$, initially at temperatures $T_A$ and $T_B$, the probability
$P(q)$ of a \emph{sharp} exchange of energy $q$ from $B$ to $A$ obeys
the relation
\begin{eqnarray}\label{JarXFT}
\frac{P(q)}{P(-q)} = \exp [\Delta \beta q ],
\end{eqnarray}
where $\Delta \beta = (kT_A)^{-1}-(kT_B)^{-1}$. This relation quantifies
the relative likelihood of a fixed exchange process and its time-reversed twin, 
and shows that heat flow from a colder to a hotter object
is exponentially suppressed. A simple application of Jensen's inequality
to (\ref{JarXFT}) leads to an averaged inequality $\<q\> (1/T_A -
1/T_B)\ge 0$. This seems to suggest that (\ref{clausius}) automatically
follows from (\ref{JarXFT}), however it must be emphasized that while
$Q_A$ equals $\<q\>$ for \emph{classical} states, this need not be true
for more general quantum mechanical states. In general, a sharp 
(rank-1 projective) energy measurement will produce 
non-classical disturbances in the quantum state of a system, and even
though the expression $\<q\>( 1/T_A - 1/T_B) \ge 0 $ still provides a
thermodynamic directionality, it is no longer identical to
(\ref{clausius}), which does not involve any measurement disturbance.


\subsection{\label{sec_Assumps}The assumption of molecular chaos and the role of correlations.}
The scope of the XFT is extremely broad, being valid for arbitrary unitary interactions between $A$ and $B$ that conserve energy, with the resultant form relying on two key assumptions of 
\begin{enumerate}[(I)]
\item initial Gibbsian states, and 
\item the assumed time-reversal invariance of the underlying dynamics.
\end{enumerate}

However the strict \emph{directionality} of the thermodynamic heat-flow relies on a third assumption 
\begin{enumerate}[(I)]
\setcounter{enumi}{2}
\item that the systems involved are initially uncorrelated - namely Boltzmann's assumption of `molecular chaos' \cite{boltzmann}.
\end{enumerate}

The molecular chaos assumption is required both in classical and quantum mechanics and, irrespective of any inherent quantum randomness,  plays the central role in thermodynamic directionality of heat flow. Indeed it has been shown explicitly that if you drop the assumption of molecular chaos then you weaken the thermodynamic arrow (as has been shown in \cite{partovi, JR1} and references therein). 

As such, both the Clausius relation, equation (\ref{clausius}), and the Jarzynski-W\'{o}jcik Fluctuation Theorem, equation (\ref{JarXFT}), are therefore limited in application, and will fail to hold within the domain of high-correlation environments. Indeed, with the extremal case of a globally pure, multipartite quantum state with thermal subsystems there should exist \emph{no directional constraint} whatsoever, and for such situations no equality such as equation (\ref{JarXFT}) should hold. The reason for this is that pure quantum states are states of maximal knowledge and may be reversibly interconverted through the appropriate unitary transformations. Such pure state thermality is of importance and turns out to be the \emph{typical} scenario with respect to the Haar measure. Specifically, for  a randomly chosen multipartite state  $|\Psi\>$ of a system with fixed energy, the reduced state for a small subsystem is exponentially likely to be Gibbsian \cite{popescushortwinter}, with the thermality arising due to quantum entanglement.

Quantum correlations can be far stronger than their classical
counterparts, and in addition to the Gibbsian typicality in pure states,
entanglement theory has other deep connections with thermodynamics
\cite{fernando2ndlaw}, often through their parallel formulations as
resource  theories \cite{resourceformulation}. Beyond the foundational
interest of studying the dissolution of the thermodynamic arrow due to
strong correlations, there is also rapid experimental progress in the
precise manipulation of small quantum systems designed to function as
engines at nanoscales, and as such, it is also of practical importance
to determine the fundamental limitations and behavior of heat exchange
in such quantum systems.

The purpose of this paper is to remove the third fundamental assumption
(III) of molecular chaos, and extend the existing XFT into
high-correlation environments, in which initially correlated quantum
systems are allowed to evolve under non-equilibrium dynamics and exchange
heat. In doing so we identify the appropriate thermodynamic
measure for the effect of correlations on sharp energy exchanges and
describe how it may contribute in work-extraction primitives, such as the maximum work theorem scenario. 

The statistics of energy exchange (and even particle exchange) has been investigated previously in a number of noteworthy publications \cite{deffner,parrondo,campisi,horowitz}. However, the important difference between these prior approaches and the work herein is the absence of initial correlations. As far as we are aware,  assumption (III) of molecular chaos has always been made, and it is not clear that previous approaches may be easily extended to this broader framework.

This paper is structured in the following way. In section \ref{generalized-setting} we present an overview of the components required for deriving the XFT, and highlight some conceptual points that relate to thermality due to quantum fluctuations. In section \ref{XFT} we analyse the heat flow in a parallel manner to the original XFT \cite{JarzynskiXFT}, however without assuming molecular chaos. We find that dropping this assumption enforces the use of a sharp mutual information measure,  quantifying the correlations between the two subsystems and how these correlations impact the thermodynamic heat flow. This in turn provides us with a generalized form of the XFT, valid in a correlated quantum environment and allowing for quantum fluctuations stemming from intrinsic randomness in pure quantum states. 
In section \ref{sec_AltXFT} we present a derivation for a non-equilibrium equality in the presence of correlations. The approach is from a more abstract setting using probability theory and random variables. The upshot of this is the acquirement of the ``efficacy factor'' that gives a notion of the strength of correlations and their effect on the directionality of thermodynamics processes. Section \ref{sec_Toy} brings all these ideas together through an exactly solvable toy model: a qubit-qudit system with an energy conserving interaction.
In section \ref{beyond-sharp} we draw attention to the strong classicality of the XFT, due to the demand of projective measurements, and highlight the technical obstacles to an extension involving more gentle, generalized POVMs. Section \ref{work-theorem} provides a simple application of our results to a maximum work theorem, and illustrates the potential work value that correlations between two thermal systems can possess.

\section{Overview of the generalized setting}\label{generalized-setting}

In this section we give an overview of the basic setting employed, including the form of our time-reversal assumptions.
Our central goal will be to establish an exchange fluctuation theorem for a bipartite
quantum system, whose reduced states $\rho_A$ and $\rho_B$ are initially
thermal, but where we drop the assumption of the initial factorization
of the joint state $\rho_{AB}$ and allow genuine quantum mechanical
coherence and entanglement to either evolve or be initially present.

\subsection{The thermodynamic scenario}
In defining heat exchange in \cite{JarzynskiXFT}, the isolated bipartite
system is assumed to undergo a three step process. An initial
energy measurement $\M_1$ is first performed on the two subsystems which
are then allowed to subsequently interact and evolve under a unitary
$U$, until a final energy measurement $\M_2$ is
performed. Thus, the bipartite quantum state
$\rho_{AB}$ undergoes the following sequence of quantum operations: 
$\rho_{AB}
\rightarrow \M_1 [\rho_{AB} ] \rightarrow U \circ \M_1[ \rho_{AB}]
\rightarrow \M_2 \circ U \circ \M_1[\rho_{AB}]$.

As mentioned, the central assumptions of the XFT are time-reversal
symmetry of the underlying dynamics and the initial thermality of the
individual subsystems.  Specifically, we assume that $\Theta^\dagger H_A
\Theta = H_A$ and $\Theta ^\dagger H_B \Theta =H_B$, where $\Theta$ is
the anti-unitary time-reversal operator. In what follows we use joint
energy eigenstates $|\phi,\chi \>$, such that $H_A |\phi,\chi \> =E_\phi
|\phi,\chi \> $ and $H_B |\phi,\chi \> =E_\chi |\phi,\chi \> $, for
which we deduce that $H_A \Theta |\phi,\chi\> = E_\phi \Theta |\phi,\chi
\>$ (with a similar expression for $B$). We take $\{|\phi\> \}$ and
$\{|\chi \> \}$ to be complete orthonormal bases for $A$ and $B$ so that
$H_A = \sum_\phi E_\phi |\phi \>\<\phi |$ and $H_B = \sum_\chi
E_\chi|\chi\>\<\chi |$, and since $\Theta$ is a symmetry of classical
states we assume that $\Theta(|\phi\>\otimes | \chi\>)$ is always in the
basis set $\{|\phi\> \otimes | \chi\>\}$ for any $\phi$ and $\chi$. The
thermal marginal states of $\rho_{AB}$ are then given by 
\begin{align}
\label{eq_therm_margA}
\rho_A &=Z^{-1}_A\sum_\phi e^{-\beta_A E_\phi} |\phi\>\<\phi|, \\
\label{eq_therm_margB}
\rho_B &= Z^{-1}_B \sum_\chi e^{-\beta_B E_\chi} |\chi\>\<\chi|, 
\end{align}
where $Z_A$ and
$Z_B$ are the usual partition functions for $A$ and $B$, and $\beta_A, \beta_B$
are their inverse temperatures.

The measurements $\M_1$ and $\M_2$ used to determine the energies of the
subsystems can in general be POVM measurements, however in what follows we restrict both
the initial and final measurement to be rank-1 projective
measurements onto the local energy eigenbases of $A$ and $B$, namely $\{|\phi\> \}$ and
$\{|\chi \> \}$, and only at the end discuss the challenges of extending beyond such sharp measurements. 

Given this setting, it is useful to introduce the notion of a \emph{history} for the composite quantum system $AB$. A history is
denoted
\begin{eqnarray}\label{history}
\gamma &=& \left (\rho_{AB} ; |\phi\> \otimes |\chi\>
	    \stackrel{U}{\longrightarrow} |\phi'\> \otimes |\chi'\>
\right ),
\end{eqnarray}
where $\rho_{AB}$ is the initial quantum state that is first projected
into the energy eigenstate $|\phi\>\otimes |\chi \>$ under $\M_1$ and
then evolves unitarily to $U (|\phi\> \otimes |\chi \>) $, which is then
measured and projected into the energy eigenstate $|\phi'\>\otimes |\chi'
\>$ under $\M_2$.

We denote by $\Gamma$ the full set of all histories $\{\gamma \} $ comprised of first
beginning in the state $\rho_{AB}$, measuring out some energy
eigenstate, evolving under some $U$ and then measuring out some final
energy eigenstate.

The thermodynamic condition of energy conservation we use is simply that
$\Tr[ \rho_{AB} (H_A+H_B)] = \Tr[U \rho_{AB} U^\dagger (H_A+H_B)]$. We
note that $U$ involves the interaction Hamiltonian that is assumed to be
smoothly switched on and off, but the energies of the subsystems are
always measured with respect to their appropriate free Hamiltonian.

\subsection{Intrinsic fluctuations due to quantum coherence}
While the exchange fluctuation theorem is a refinement on the
thermodynamic arrow for heat flow, in that it deals with subensembles of
postselected outcomes with sharp energy transfers $q$, not all of
the uncertainty is attributable to the statistical mixture of different
energy states. Pure quantum mechanical states allow the possibility of
intrinsic quantum fluctuations, and so while we might step
beyond classical statistical fluctuations by focusing on individual
pure state outcomes in the XFT, we might also allow the possibility of
quantum coherence evolving under the unitary dynamics and generating new
indeterminacy. For example, with respect to the average statistics of energy
measurements $\{|E_k\>\<E_k|\}$, the pure quantum state $|\psi\> \propto \sum_k
\sqrt{e^{-\beta E_k}} |E_k\>$ is indistinguishable from a thermal mixed
state $\rho_{\mathrm{therm}}=Z^{-1}\sum_k e^{-\beta E_k} |E_k\>\<E_k|$.

Nevertheless, the energy measurement of such superpositions $|\psi\> $ will display
quantum fluctuations, some of which may increase the total energy of
$AB$, some decrease it, but on average no net energy should be gained
from the fluctuations. It is then simple to allow histories with
positive energy fluctuations that increase the total energy of $AB$, or
negative fluctuations that decrease the energy of $AB$, and also
histories with no fluctuations at all. As such, a useful and physically
intuitive division of the set of histories is into sets of histories
with similar energy transformations. In particular we write $\Gamma =
\cup_{q, \Delta \epsilon} \Gamma(q, \Delta \epsilon)$, where $\Gamma
(q, \Delta \epsilon)$ is the set of $\gamma$ (of the form
(\ref{history})) with fluctuations of the total energy of $AB$
\begin{align}
\Delta \epsilon = \<\phi', \chi'| (H_A+H_B) |\phi',\chi'\> -\<\phi,
 \chi| (H_A+H_B) |\phi,\chi\>
\end{align}
and with an energy transfer $q$ into $A$ defined as 
\begin{equation}
q = \<\phi '|H_A |\phi ' \> - \<\phi  |H_A |\phi  \>.
\end{equation}

In the next section we make use of this setting and derive the generalized XFT theorem.

\section{An Exchange Fluctuation Theorem for correlated quantum states}\label{XFT}
We can now formulate a generalized XFT that drops assumption (III) of molecular chaos (see section \ref{sec_Assumps}) and allow a general bipartite quantum state $\rho_{AB}$ with thermal marginals.  Given this initial state $\rho_{AB}$, the occurrence of a single history
$\gamma \in \Gamma( q, \Delta \epsilon)$ in equation \eqref{history}, has probability
\begin{eqnarray}\label{eq_Pgamma}
\mbox{Prob} [\gamma ] = \<\phi,\chi | \rho_{AB} |\phi,\chi \> |
 (|\phi',\chi' \> , U |\phi,\chi\>)|^2,
\end{eqnarray}
where 
$U \equiv e^{-iHt}$
and 
$H \equiv H_{A} + H_{B}+ H_{\rm int}$ 
is the total Hamiltonian, including the interaction
$H_{\rm int}$ 
between $A$ and $B$, 
which is switched on at the initial time, 
and for clarity we write the Hilbert space inner product
as $ ( \cdot , \cdot )$.

From time-reversal invariance, $\Theta^\dagger H \Theta = H$, and the anti-unitarity of $\Theta$ it
follows that $U =\Theta^\dagger U^\dagger \Theta$,  and so we have that
\begin{eqnarray}
\label{dynamical}
|(|\phi',\chi' \> , U|\phi,\chi\>)|^2 &=&|(\Theta |\phi,\chi \> , U \Theta |\phi',\chi' \> )|^2.
\end{eqnarray}
Thus, time-reversal symmetry alone implies the probability to go from
the initial state $|\phi,\chi\>$ to $|\phi',\chi' \>$ is always equal to
the probability to go from $\Theta |\phi',\chi'\>$ to $\Theta |\phi,\chi
\>$ under the same unitary interaction. In what follows, we use a star
to denote time-reversed objects, for example $|\phi', \chi ' \>_* :=
\Theta |\phi' , \chi '\>$.

To quantify correlations in the quantum state as they relate to the XFT
we define, for any joint local POVMs $\{M_i\}$ on $A$ and $\{N_j\}$ 
on $B$, the quantity  $\II(\rho_{AB}; M_i, N_j)$ via the expression
\begin{eqnarray}
\II(\rho_{AB}; M_i, N_j) &:=& \ln \left ( \frac{\Tr[ M_i \otimes N_j\rho_{AB}]}{\Tr[M_i
\rho_{A}]\Tr[N_j \rho_{B}]} \right ) .
\end{eqnarray}


For a sharp energy measurement, we simply have $\M_1 = \{M_\phi
\otimes N_\chi \}$ where $\{ M_\phi=|\phi\>\<\phi| \}$ and
$\{N_\chi=|\chi \>\<\chi| \}$ are the rank-1 projectors in the
energy eigenbases\footnote{The function $\II$ may be related to the
classical relative entropy of the joint measurement outcomes through the
relation
$I_c(\M:\N) =\sum_{i,j} \Tr[ M_i\otimes N_j \rho_{AB}] \II(\rho_{AB}; M_i, N_j)$.},
and when $\rho_{AB}$ is a correlated quantum state having thermal
marginals as in equations \eqref{eq_therm_margA} and \eqref{eq_therm_margB},
we find that $\M_1$ maps $\rho_{AB}$ into a classically
correlated state, diagonal in the energy eigenbasis. Moreover, it is
readily seen that $\Tr_{B (A)} [ \M_1 [ \rho_{AB}] ]= \rho_{A
 (B)}$, and so the state $\M_1 [ \rho_{AB}]$ has the same thermal
marginals as $\rho_{AB}$ \footnote{In addition, we have that $I_c
(\M:\N)= I[ \M_1[\rho_{AB}]
; A:B]$ where $I[ \sigma_{AB} ; A:B]$ is the quantum mutual information
of the bipartite state $\sigma_{AB}$, defined as $I[ \sigma_{AB} ; A:B]
= S[\sigma_A] + S[\sigma_B] - S[\sigma_{AB}]$. }.

For this particular initial measurement, the probability
$p(|\phi,\chi\>)$ of projecting into the state $|\phi, \chi\>$ under
$\M_1$ can be written as
\begin{eqnarray}
\hspace{-0.5cm}
p(|\phi,\chi \>) &=& e^{-\beta_A E_\phi -\beta_B E_\chi - \log(Z_AZ_B) +
 \II[\rho_{AB}; M_\phi, N_\chi]},
\end{eqnarray}
while a comparison with the probability of obtaining the $|\phi', \chi ' \>_*$ outcome implies that
\begin{eqnarray}
\label{thermal}
p(|\phi,\chi\>) &=& p(|\phi',\chi'\>_*) e^{\Delta \beta q +\beta_B \Delta \epsilon - \Delta \II(\gamma)},
\end{eqnarray}
where $\Delta \beta = \beta_A -\beta_B$, and crucially 
\begin{equation}
\Delta \II
(\gamma) = \II[\rho_{AB};M_{\phi_*'}, N_{\chi_*'} ] - \II[\rho_{AB};
M_\phi,N_\chi]
\end{equation}
is the appropriate correlation measure, dependent only
on the initial state $\rho_{AB}$ and the initial measurement $\M_1$. A derivation of (\ref{thermal}) is provided in the appendix. Note that with this definition of $\Delta \II(\gamma) $, the assumption of molecular chaos gives $\Delta \II
(\gamma) =0$.

Combining (\ref{thermal}) with (\ref{dynamical})  we find that
\begin{eqnarray}
\label{singlehist}
\frac{\rm{Prob}[\gamma]}{\rm{Prob}[\gamma^*]} &=& e^{\Delta \beta q +\beta_B \Delta \epsilon - \Delta \II(\gamma)},
\end{eqnarray}
where $ \gamma^* $ is the time-reversed twin of $ \gamma$ given by
\begin{eqnarray}
\gamma^* \equiv (\rho_{AB} ; |\phi',\chi'\>_* \stackrel{U}{\longrightarrow} |\phi, \chi\>_*).
\end{eqnarray}
In particular, the history $\gamma$ involves a quantity of energy $q$ being
transferred into $A$ and a net increase of total energy $\Delta
\epsilon$, while $\gamma^*$ involves the opposite changes, and so
$\Gamma(q, \Delta \epsilon)^* = \Gamma (-q , -\Delta \epsilon)$. We also
note that (\ref{singlehist}) is independent of the specific form of the
dynamics (beyond time-reversal invariance), and depends solely on the
properties of the initial quantum state $\rho_{AB}$.

One can now compare the ratio of probabilities of the set $\Gamma(q,\Delta \epsilon )$ and its time-reversed twin set $\Gamma (q, \Delta \epsilon)^*= \Gamma (-q, -\Delta \epsilon)$. The probability of the former is given by
\begin{align}
\hspace{-0.4cm}\mathrm{Prob} [\Gamma(q, \Delta \epsilon)] &=& \hspace{-0.4cm}\sum_{\gamma \in \Gamma (q,\Delta \epsilon) } \hspace{-0.2cm}e^{\Delta \beta q +\beta_B \Delta \epsilon - \Delta \II(\gamma)} \mathrm{Prob}[\gamma^*].
\end{align}
While the term $e^{\Delta \beta q+\beta_B \Delta \epsilon} $ may be
factored out of the sum, the correlation term cannot as it will
generally vary over the set $\Gamma (q,\Delta \epsilon)$. Instead we
necessarily obtain bounds for the ratio of the probabilities. To fix the
lower and upper bounds, we respectively define $ \Delta \II_l =
\mbox{max}_{\gamma \in \Gamma (q,\Delta \epsilon)} [\Delta \II(\gamma)
]$ and $ \Delta \II_u = \mbox{min}_{\gamma \in \Gamma (q,\Delta
\epsilon)} [\Delta \II(\gamma) ]$, and immediately deduce that
\begin{align}\label{genXFT}
\hspace{-0.2cm} e^{\Delta \beta q +\beta_B\Delta \epsilon -\Delta \II_l} \le \frac{\mathrm{Prob}[\Gamma(q,\Delta \epsilon)]}{\mathrm{Prob}[\Gamma(-q, -\Delta \epsilon)]} \le e^{\Delta \beta q +\beta_B \Delta \epsilon-\Delta \II_u}.
\end{align}

The XFT in equation (\ref{genXFT}) is a
generalization of Jarzynski and W\'ojczik in equation (\ref{JarXFT}) and it
is a constraint on the relative likelihood of a forward transition to a
backward transition given an initially correlated quantum state
with thermal subsystems. As can be seen, 
by moving away from the assumption of molecular chaos we obtain $\Delta \II \neq 0$ and
one gradually weakens the
constraint on the thermodynamic arrow, as expected. Moreover, it is not
possible to tighten these bounds without making additional assumptions
as to the particular form of the dynamics.

Beyond the relative likelihood of the forward and reverse processes, one
can take (\ref{singlehist}) and sum over $\gamma \in \Gamma$, to obtain
the non-equilibrium equality for an initially correlated state
\begin{eqnarray}\label{genJarzynski}
\<e^{-\Delta \beta q - \beta_B \Delta \epsilon + \Delta \II}\> = 1
\end{eqnarray}
and then using Jensen's inequality we have that
\begin{eqnarray}\label{average}
\Delta \beta \< q\> + \beta_B \< \Delta \epsilon\> - \< \Delta \II \> \ge 0.
\end{eqnarray}
Here, $\< \Delta \II \>$ represents the difference in the
classical mutual information of measurement outcomes between the initial
and final states.

Given the assumption that $\< \Delta \epsilon \>=0$ we have $\Delta
\beta \<q\> - \<\Delta \II \> \ge 0$, which reduces to \eqref{clausius} the Clausius
relation $Q(1/T_A - 1/T_B)\ge 0$ for $Q=\<q\>$ and the assumption of
molecular chaos. More importantly it displays the energetic value of
correlations in providing a modified lower bound of $Q(1/T_A-1/T_B) \ge
\< \Delta \II \>$ with the function $\II(\rho_{AB} ; M_i, N_j)$ as the
appropriate sharp-outcome measure for the initial bipartite quantum
state. This must be compared with averaged results obtained previously
\cite{partovi, JR1} in which $\<\Delta \II \>$ is replaced with the
change in the quantum mutual information of the state $\rho_{AB}$. The
origin of the difference is that the XFT demands sharp energies at the
initial and final stages, as opposed to bluntly looking at expectation
values of energy for pure quantum states.

While it is natural to impose energy conservation, either at the level
of commuting Hamiltonians or expectation values, the XFT given by
equation (\ref{genXFT}) makes predictions for a particular type of state
with local temperatures and global correlations, and provides only the
relative likelihood of seeing one forward thermodynamic process compared
to its reverse - not whether it occurs at all. As such, the specific
interaction Hamiltonian that is used only serves to predict the absolute
likelihood of these different processes.

\section{\label{sec_AltXFT}A non-equilibrium equality in the presence of correlations}

In the previous section we derived a non-equilibrium equality, equation \eqref{genJarzynski}, for initially correlated systems using the concrete idea of histories and time reversal. Here, we follow the compact approach of \cite{Lidar} to isolate the ``correlation factor'' that quantifies the deviation from molecular chaos.

We adopt the same prepare-evolve-measure setting as before. The initial bipartite quantum state $\rho$ is projected onto the energy basis $\mathcal{M}_1=\{M_\phi \otimes N_\chi\} = \{\ket{\phi}\bra{\phi}\otimes \ket{\chi}\bra{\chi}\}$. For simplicity we use $\mu = (\phi,\chi)$ to label the outcome of $\mathcal{M}_1$ that prepares the state $$\rho_\mu = \frac{1}{p_\mu}M_\mu \otimes N_\mu \rho M^\dag_\mu \otimes N^\dag_\mu = \ket{\phi, \chi}\bra{\phi, \chi}$$ with probability
\begin{align}
p_\mu = \Tr[M_\mu \otimes N_\mu \rho M^\dag_\mu \otimes N^\dag_\mu ].
\label{p_mu}
\end{align}
The initial state $\rho$ is again assumed to have thermal marginals from which we define the uncorrelated probability distribution
\begin{align}
p^0_\mu = \Tr[M_\mu \rho_A] \Tr [ N_\mu \rho_B]
=\frac{e^{-\beta_A \phi}}{Z_A}\frac{e^{-\beta_B \chi}}{Z_B}.
\end{align}

The prepared state $\rho_\mu$ evolves under the unitary $U$ to $\rho'_\mu = U \rho_\mu U^\dag$ and after the interaction the final energy measurement projects this state onto $\rho'_{\nu|\mu} = \frac{1}{p_{\nu|\mu}}M_\nu \otimes N_\nu \rho_\mu M_\nu \otimes N_\nu = \ket{\phi',\chi'}\bra{\phi',\chi'}$ with the outcome labelled by $\nu=(\phi',\chi')$ and probability 
\begin{align}
\label{p_cond}
p_{\nu|\mu} &= \Tr[M_\nu \otimes N_\nu \rho'_\mu M^\dag_\nu \otimes N^\dag_\nu ]= |\langle \phi',\chi' | U | \phi,\chi \rangle |^2.
\end{align}
The total probability to obtain outcome $\nu$ 
is $p_\nu = \sum_\mu p_\mu p_{\nu|\mu} := \sum_\mu p_{\mu\nu} $, where
\begin{align}
\label{eq_pmunu}
p_{\mu\nu} = \bra{\phi,\chi}\rho\ket{\phi,\chi} |\langle \phi',\chi' | U | \phi,\chi \rangle |^2
\end{align}
is simply $\mathrm{Prob}[\gamma]$ in equation \eqref{eq_Pgamma}.

We convert $p_{\mu\nu}$ into a probability density function on $\mathbb{R}$ for a continuous random variable $x$ by writing
\begin{equation}
P_X(x) = \sum_{\mu\nu} \delta(x-X_{\mu\nu}) p_{\mu\nu} 
\end{equation}
where $X_{\mu\nu} $ is a discrete random variable distributed according to $p_{\mu\nu} $. 
Define the function $F_{\wt X}(-x) = P_X(x)e^{-x}$ that is analogous to the time-reversed probability in the XFT. It may be shown that $F_{\wt X}$ is a probability density function for the random variable $\wt X := -X$ if $p_{\mu\nu} e^{-X_{\mu\nu}}$ is a probability distribution (see the Appendix for details).

We now choose the random variable $X_{\mu\nu}$ to be given by
\begin{equation}
X_{\mu\nu} = \ln p_\mu - \ln f_\nu + \Delta \II_{\mu\nu}.
\end{equation}
where the correlation $$\Delta \II_{\mu\nu} = \II_\nu - \II_\mu=\ln \left(\frac{p_\mu}{p^0_\mu} \right)-\ln \left(\frac{f_\nu}{f^0_\nu}\right)$$ and $f_\nu=\Tr[M_\nu \otimes N_\nu \rho M^\dag_\nu \otimes N^\dag_\nu ]$, $f^0_\nu=\frac{e^{-\beta_A \phi'}}{Z_A}\frac{e^{-\beta_B \chi'}}{Z_B}$ are the probabilities of the final measurement on the correlated $\rho$ and product $\rho_A \otimes \rho_B$ states.

By taking an average of both sides of $F_{\wt X} = P_Xe^{-x}$, this choice of variables gives us the thermodynamic relation
\begin{equation}
\label{eq_alt_NEeq}
\langle e^{-\Delta \beta q - \beta_B \Delta \epsilon} \rangle_p = \langle e^{-\Delta \II} \rangle_f.
\end{equation}
The subscripts $p$ and $f$ indicate that the averages are to be taken with respect to the probability distributions $p_{\mu\nu}$ (equation \eqref{eq_pmunu}) and
\begin{align}
f_{\mu\nu} &=  \bra{\phi',\chi'}\rho\ket{\phi',\chi'} |\langle \phi,\chi | U^\dag | \phi',\chi' \rangle |^2.
\end{align}
The sharp heat into A is $q= \phi' - \phi$, the inverse temperature difference $\Delta \beta = \beta_A -\beta_B$, and the global energy change $\Delta \epsilon =\phi' + \chi'- \phi - \chi$. This formulation separates out the ``correlation factor''
$\eta := \langle e^{-\Delta \II} \rangle_f $ that quantifies the deviation from the assumption of molecular chaos.
In comparing equations \eqref{eq_alt_NEeq} and \eqref{genJarzynski} we notice that the above analysis implies that ``taking $\Delta \II$ to the other side'' in \eqref{genJarzynski} results in now having to take the average with respect to $f_{\mu\nu}$, the time-reversed probability distribution $p_{\mu\nu}$. A further difference between equations \eqref{genJarzynski} and \eqref{eq_alt_NEeq} is that applying Jensen's inequality here gives $$\Delta\beta\langle q\rangle + \beta_B \langle \Delta \epsilon \rangle + \ln\eta \geq 0.$$ This matches equation \eqref{average} if and only if the random variable $\Delta \II = \mathrm{const}$.

\section{\label{sec_Toy}An exactly solvable toy model}

We now compare three ``lenses'' through which to view heat exchange between correlated systems: the XFT from equation \eqref{genXFT}, its averaged form in equation \eqref{average}, and the exponentiated average in \eqref{eq_alt_NEeq}. For convenience we list them below in order
\begin{subequations}
\begin{align}
\hspace{-0.2cm} e^{\Delta \beta q +\beta_B\Delta \epsilon -\Delta \II_l} &\le \frac{\mathrm{Prob}[\Gamma(q,\Delta \epsilon)]}{\mathrm{Prob}[\Gamma(-q, -\Delta \epsilon)]} \le e^{\Delta \beta q +\beta_B \Delta \epsilon-\Delta \II_u}, \\
&\Delta \beta \< q\> + \beta_B \< \Delta \epsilon\> - \< \Delta \II \> \ge 0, \\
&\langle e^{-\Delta \beta q - \beta_B \Delta \epsilon} \rangle_p = \langle e^{-\Delta \II} \rangle_f=:\eta.
\end{align}
\end{subequations}
To interpret these relations, we consider a setting that admits a complete solution. Significant differences arise between them even for a low-dimensional scenario, for which we have the usual caveat that distributions can become quite broad, and so any expectation values must correspond to multiple runs on the systems in the i.i.d. limit.

We work with a joint Hilbert space $\H_{d_A} \otimes \H_{d_B}$ describing two subsystems of dimension $d_A=2$ and $d_B=d$, with $d$ unspecified for now. 
The free Hamiltonian of system $i \in \{ A,B \}$ is
\begin{equation}
H_i = \sum_{n=0}^{d_i-1} n\ket{n}\bra{n}
\end{equation}
where we have set all energy separations to be unity and the ground state is zero. 

The energy-conserving interaction Hamiltonian that commutes with the free part $H_A + H_B$ is
\begin{align}
 H_{\rm int} = \sum_{j=1}^{d-1} \omega_j (\ket{0,j}\bra{1,j-1}+\ket{1,j-1}\bra{0,j})
\end{align}
The eigenvectors of $ H_{\rm int}$ are $\ket{\pm_j} = \frac{1}{\sqrt{2}}(\ket{1,j-1}\pm\ket{0,j})$ with eigenvalues $\pm \omega_j$, for $j=1,\dots,d-1$.
Coherent evolution under this Hamiltonian is restricted to the energy-degenerate, two-dimensional subspaces spanned by $\{ \ket{0,j},\ket{1,j-1}\}$.

In deriving the thermodynamic relations, an initial projective measurement $\mathcal{M}_1$ is made on the bipartite state $\rho$, this kills off any coherence in the free energy eigenbasis. As such we simply take as our initial state a classically correlated density matrix 
\begin{equation}
\rho_{AB} \equiv \rho_{AB}(t=0) = \sum_{m=0}^{1}\sum_{n=0}^{d-1} \lambda_{mn} \ket{mn}\bra{mn}
\end{equation}
where $\sum_{mn} \lambda_{mn} = 1$ and $\lambda_{mn} \geq 0$ for all $m,n$.
A further constraint on the $\lambda_{mn}$ comes from the requirement that the subsystems $i = \{A,B\}$ must be thermal states $\Tr_{\backslash i}[\rho_{AB}]  = \frac{1}{Z_i} e^{-\beta_i H_i},$ the $\backslash i$ notation means the complement of $i$. 
The bipartite state evolves to $\rho_{AB}(t) = U(t) \rho_{AB} U(t)^\dag$, where $U(t) = e^{-iHt}$, with $H = H_A + H_B + H_{\rm int}$ and then a final measurement is made in the free energy basis.

In this toy example the three thermodynamic relations become
\begin{subequations}
\begin{align}
\label{eg_XFT}
&&\min\left\{\frac{\lambda_{0,j}}{\lambda_{1,j-1}}\right\} \leq \frac{\mathrm{Prob}[\Gamma(q=1)]}{\mathrm{Prob}[\Gamma(q=-1)]} \leq \max\left\{\frac{\lambda_{0,j}}{\lambda_{1,j-1}}\right\} \\
&&\Delta\beta \< q \> \geq \sum_j (\lambda_{0j}-\lambda_{1,j-1})\left( \Delta \beta + \ln \frac{\lambda_{1,j-1}}{\lambda_{0,j}}\right)\sin^2\omega_jt \\
\label{simp_eta}
&&\eta = 1+ \sum_{j=1}^{d-1} (\lambda_{0j}(e^{-\Delta \beta}-1)+ \lambda_{1,j-1}(e^{\Delta \beta}-1))\sin^2\omega_j t
\end{align}
\end{subequations}
We refer the reader to the Appendix for details.

To analyse these results, we concentrate on the smallest $d$ that gives non-trivial results. Note that the condition that $\rho_{AB}$ is physical has not been imposed yet. The initial density matrix is required to have thermal marginals. A convenient way of describing the correlated bipartite state $\rho_{AB}$ on $\H_{d_A} \otimes \H_{d_B}$ is by writing it as $\rho_{AB} = \rho_A \otimes \rho_B + \tau_{AB}$, where the operator $\tau_{AB}$ must obey $\Tr_A [\tau_{AB}] = 0$ and $\Tr_B[ \tau_{AB}] =0$ to ensure that $\rho_{AB}$ has thermal marginals. Furthermore, we must have that $\Tr [ \tau_{AB}] = 0$ and $\rho_A \otimes \rho_B + \tau_{AB} \ge 0$ to ensure that $\rho_{AB}$ is a genuine quantum state. Let $\tau_{AB}$ be a diagonal matrix, initially it has $2d$ parameters, but there are three constraints therefore we reduce to $2d-3$ independent parameters, and in fact we always look for the smallest number of independent parameters. For less cumbersome notation, define $\zeta := (Z_A Z_B)^{-1}$, $a := \beta_A$ and $b:=\beta_B$ so that $\Delta\beta = a - b$. Also in the following we set $\omega_j = \omega$ for all $j$ and analyse the systems at the time where $\omega t = \pi/2$.

\subsection{A two-qubit system: $d=2$}

The matrix $\tau_{AB} = \zeta\diag(x,-x,-x,x)$ satisfies the trace conditions on $\tau_{AB}$ for some $x \in \mathbb{R}$. Positivity of the matrix $\rho_{AB} = \rho_A \otimes \rho_B + \tau_{AB}$ leads to the constraint
\begin{align}
-e^{-(a+b)}\leq x \leq \min\{e^{-a},e^{-b}\}.
\end{align}
Let's take $B$ to be hotter than $A$ at the start, then $\Delta\beta > 0$ and $\min\{e^{-a},e^{-b}\} = e^{-a}$. Notice that high temperatures (i.e. small $a,b$) widen the range of $x$ because larger temperatures are synonymous with reduced states being more mixed and this permits greater correlations. 
Transitions occur only in between the energy degenerate states $\ket{01}$ and $\ket{10}$ and we have $\lambda_{01}=\zeta(e^{-b}-x)$ and $\lambda_{10}= \zeta(e^{-a}-x)$.

The three thermodynamics relations become
\begin{eqnarray*}
&&\frac{\mathrm{Prob}[\Gamma(q=1)]}{\mathrm{Prob}[\Gamma(q=-1)]} = \frac{e^{-b}-x}{ e^{-a}-x},\\
&&\Delta\beta \< q \> \geq \zeta(e^{-b}-e^{-a})\left(\Delta\beta+ \ln \frac{e^{-a}-x}{e^{-b}-x}\right),\\
&&\eta = 1 + \zeta( (e^{-b}-x)(e^{-\Delta \beta}-1)+ (e^{-a}-x)(e^{\Delta \beta}-1)). 
\end{eqnarray*}
We analyse these relations in different extremal settings.

\bigskip

\textit{Equal temperature $\Delta\beta=0$:}
When the qubits are at equal temperatures, the likelihoods of the forward (heat from $B$ to $A$: $q>0$) to backward (heat from $A$ to $B$: $q<0$) transitions are equal $\mathrm{Prob}[\Gamma(q=1)]=\mathrm{Prob}[\Gamma(q=-1)]$. Thus \emph{detailed balance} is preserved no matter the size of initial correlations.

\textit{Maximum value of $x = e^{-a}$ with $\Delta\beta >0$:}
When $x$ takes this value, the $\lambda_{10}$ eigenvalue is set to zero. Since the interaction is between the $\ket{01}$ and $\ket{10}$ states only, switching off $\lambda_{10}$ means that the only transition allowed is the forward one: $\ket{01} \rightarrow \ket{10}$. This is a \emph{deterministic} transfer of heat from $B$ to $A$ and is possible due to the correlations, and regardless of the initial temperatures as long as $T_B > T_A$. This behaviour is reflected in the thermodynamic relations. Let $$R := \frac{\mathrm{Prob}[\Gamma(q=1)]}{\mathrm{Prob}[\Gamma(q=-1)]}$$ then
in this case, the ratio diverges $R \rightarrow \infty$ as expected since the backward transition (the denominator) does not occur $\mathrm{Prob}[\Gamma(q=-1)]=0$. In this limit the Clausius relation trivialises $\Delta\beta \geq -\infty$. In contrast, the correlation factor remains finite $\eta = 1-\zeta  e^{-b}(1-e^{-\Delta \beta})^2 < 1$. As this is equal to $\langle e^{-\Delta\beta q}\rangle$, and by assumption $\Delta \beta > 0$, this reflects the fact that the transition probability distribution is skewed so that the $q > 0$ transition is far more likely than the backward one. Unlike the first two thermodynamic relations, the correlation factor is sensitive to the temperatures of the two qubits even in this extremal case. As long as $x=e^{-a}$ then $\lambda_{01}=0$, however $\lambda_{10}=\zeta(e^{-b}-e^{-a})$ remains temperature dependent and varies between $0$ for $a=b$ and $\frac 14$ for $b<<a<<1$, the limit where $A$ and $B$ are very hot but $A$ at significantly cooler than $B$. Since $\eta$ is linear in $\lambda_{ij}$ it varies with the the choice of $a$ and $b$ through $\lambda_{10}=\zeta(e^{-b}-e^{-a})$. We find the smallest value $\eta$ can attain is $\frac 34$ for the most skewed distribution permitted $\{\lambda_{00},\lambda_{01},\lambda_{10},\lambda_{11}\} = \{\frac 12,\frac 14,0,\frac 14\}$ which delivers the biggest amount of heat $q=1$ per bipartite system $\rho$.

\textit{Minimum value $x = -e^{-(a+b)}$ with $\Delta\beta >0$}:
We just saw that it is possible for correlations to switch off the backwards transition by setting $\lambda_{10} = 0$, however, we can never completely suppress the forward transition. At the minimum value of $x$ we have $\lambda_{01} =\zeta e^{-b}(1+e^{-a})$ and $\lambda_{10} = \zeta e^{-a}(1+e^{-b})$ giving $$R = e^{\Delta\beta}\left(\frac{1+e^{-a}}{1+e^{-b}}\right).$$
When both temperatures are low so that $e^{-a}, e^{-b} << 1$ then we approach the uncorrelated ratio $R\approx e^{\Delta\beta}$, this is a reflection on the fact that at low temperatures the reduced states are more pure and therefore cannot be highly correlated. Otherwise $R < e^{\Delta\beta}$, since $e^{-a} < e^{-b}$ by assumption, so that for maximal $x$, the ratio of forward to backward transfer can be suppressed compared to the uncorrelated case. However there is no amount of correlation that makes $R < 1$ which would mean that negative heat flow is always more likely to occur.

The Clausius relation becomes $$\Delta\beta \< q \> \geq \zeta(e^{-b}-e^{-a})\ln\left(\frac{1+e^{-b}}{1+e^{-a}}\right) \geq 0$$ 
so we see that with such correlations, we are guaranteed that a finite amount of averaged heat will transfer from hot to cold (unless $a=b$), but the probability for doing so shrinks compared to the uncorrelated case.

Finally the correlation factor $$\eta = 1+\zeta(e^{-a}-e^{-b})^2$$ is greater than unity because the forward process is reduced and we are relatively more likely to observe a sharp amount of heat $q=-1$ being transferred, even though on average $\< q \> > 0$.

Even in the elementary system of two qubits, we observe rich heat-exchange behaviour as captured by our three thermodynamic relations. The XFT is depicted in the top of figure 1. We now consider a qubit-qutrit system in which correlations lead to even more non-classical features.

\subsection{A qubit-qutrit system: $d=3$}

It is easily checked that the matrix $\tau_{AB} = \zeta\diag(x,y,-(x+y),-x,-y,x+y)$ satisfies the trace conditions on $\tau_{AB}$ for some $x,y \in \mathbb{R}$. Positivity of the matrix $\rho_{AB} = \rho_A \otimes \rho_B + \tau_{AB}$ leads to
\begin{align}
-1\leq &x \leq e^{-a}\\
-e^{-b}\leq &y \leq e^{-(a+b)}\\
-e^{-(a+2b)}\leq x&+y\leq e^{-2b}
\end{align}
As for the $d=2$ case, higher temperatures, corresponding to lower values of $a,b$, allow a greater variation of initial correlations. 


This time transitions occur within two subspaces: $\{\ket{01},\ket{10}\}$ and $\{\ket{02},\ket{11}\}$. We have 
\begin{align}
\lambda_{01}&=\zeta(e^{-b}+y)\\
\lambda_{10}&= \zeta(e^{-a}-x)\\
\lambda_{02}&=\zeta(e^{-2b}-(x+y))\\
\lambda_{11}&=\zeta(e^{-(a+b)}-y)
\end{align}

In this case the three thermodynamic relations are
\begin{widetext}
\begin{eqnarray}
&&\min\left\{\frac{e^{-b}+y}{e^{-a}-x}, \frac{e^{-2b}-(x+y)}{e^{-(a+b)}-y} \right\} \leq \frac{\mathrm{Prob}[\Gamma(q=1)]}{\mathrm{Prob}[\Gamma(q=-1)]} \geq \max\left\{\frac{e^{-b}+y}{e^{-a}-x},\frac{e^{-2b}-(x+y)}{e^{-(a+b)}-y} \right\}\\
&&\Delta\beta \< q \> \geq  \Delta\beta\zeta((1+e^{-b})\delta+y)+ \zeta\left((\delta+x+y) \ln\frac{e^{-a}-x}{e^{-b}+y}+ (e^{-b}\delta-x)\ln\frac{e^{-(a+b)}-y}{e^{-2b}-(x+y)}\right)\\
&&\eta = 1+ \zeta((e^{-b}(1+ e^{-b})-x)(e^{-\Delta \beta}-1) + (e^{-a}(1+e^{-b})-(x+y))(e^{\Delta \beta}-1))
\end{eqnarray}
\end{widetext}
and $\delta:=e^{-b}-e^{-a}$.

\textit{Equal temperature $\Delta\beta=0$:}
The situation now is entirely different to the two qubit case: at equal temperatures the ratio of probabilities $R$ in the XFT is not equal to unity hence correlations distort detailed balance (for finite $a$)! Depicted in the bottom of figure 1. There are two choices that make the upper bound diverge for 
\begin{eqnarray}
&&\frac{e^{-a}(1-e^{-a})-y}{e^{-2a}-y} \leq R < \infty, \quad x=e^{-2a}\\
&&\frac{e^{-a}(1+e^{-a})}{e^{-a}-x} \leq R < \infty, \quad \quad y=e^{-2a}
\end{eqnarray}
but it is not possible to simultaneously set $x=y=e^{-2a}$ so we cannot attain $R \rightarrow \infty$.

Consider the top relation. We tighten the range of $R$ when the lower bound is maximised, that is, fixing $y=e^{-a}(e^{-a}-1)$. Doing this we find that the lower bound is greater than unity if $T_A = T_B > (\ln 2)^{-1}$. So by setting the temperature high enough, we are more likely to observe heat flow into system A. At a lower temperature, this is not guaranteed as the lower bound can go to zero in which case $R\in [0,\infty]$ and the bounds are not tight at finite temperatures. Regarding the bottom inequality for $R$, we may maximise the lower bound by setting $x=0$ and find $R \in [1,\infty]$ meaning that that qutrit is more likely to heat the qubit.
We see that correlations make heat flow asymmetric even when $A$ and $B$ are at the same temperature. Observing a bias in the direction of heat flow when $\Delta\beta=0$ therefore provides a way of revealing the difference between system sizes.

For $\Delta\beta = 0$ the Clausius inequality is not illuminating, however the correlation factor $$\langle e^{-\Delta\II}\rangle = 1$$ is non-trivial because $\Delta\II \neq 0$ in general for the qubit-qutrit case when $\Delta\beta=0$ (the expression for $\Delta\II$ is given in the Appendix, and it is equal to zero when $\Delta\beta=0$ for the qubit-qubit system). We can interpret this expression as a ``correlation fluctuation theorem'' for systems at equal temperature.

\textit{Unequal temperature $\Delta\beta\neq 0$:}
There are choices of $x,y$ that can completely switch off one transition $\ket{0,j} \leftrightarrow \ket{1,j-1}$ thus reducing the qubit-qutrit system to an effective qubit-qubit one. For instance $y=e^{-(a+b)}$ and $x = e^{-b}(e^{-b} - e^{-a})$ sets $\lambda_{11} = \lambda_{02}  =0$ and only the $\ket{01} \leftrightarrow \ket{10}$ transition is allowed, and we still have the freedom to make it deterministic so that $\ket{01} \rightarrow \ket{10}$ only, if additionally $ e^{-b}(e^{-b} - e^{-a}) = e^{-a}$ is satisfied. Similarly choosing $y = -e^{-b}$ and $x = e^{-a}$ switches of the $\ket{01} \leftrightarrow \ket{10}$ and, if $ e^{-a}-e^{-b} = e^{-2b}$ is satisfied, then $\ket{02} \leftrightarrow \ket{11}$ is deterministic. These settings recover the qubit-qubit case where we can get deterministic heat flow from hot to cold by picking the correlations correctly.

\section{\label{sec_Phys_Disc}Discussion of physical assumptions}

Some of the assumptions and terms we have introduced require clarification and comparison with current literature on the topic of XFTs.

\subsection{Macroscopic significance of the function $\II$}
The quantity $\II$ might at first glance seem to be merely a mathematical measure of correlation without any operational significance, however this is not the case and it is simply a sharp version of the mutual information $I = \langle \II \rangle$, which in turn arises in extremely natural macroscopic and operational situations. For example, it is known to have the operational meaning as the work required to decorrelate a system in the asymptotic/macroscopic regime \cite{groisman}, while in other thermodynamic contexts it is identified as the correct measure of correlations in thermodynamic processes of bipartite quantum thermal systems \cite{partovi, JR1} for averaged measurement outcomes. Finally the role of mutual information in thermodynamics also arises in the context of Maxwell Demon scenarios \cite{Zurek}, in which the extractable work is given by $W \le kT I(X:Y)$, where $Y$ is the measurement statistics of the demon and $X$ is the actual microstate of the physical system. This energetic value of correlations can be cast into the form of a fluctuation theorem that amounts to a work extraction version of the results presented here, and recently has been experimentally verified in the context of feedback control of microscale thermodynamic systems \cite{InfoHeatEngine}.

\subsection{But shouldn't fluctuation theorems be equalities, not inequalities?}

That we have obtained an \emph{inequality} in equation \eqref{genXFT} to describe the high-correlation scenario might seem as a step in the wrong direction, given that fluctuation theorems give \emph{equalities} that generalize the more traditional inequalities such as the Clausius relation $-W \le -\Delta F$. However it is easy to see that equation (\ref{genXFT}) is indeed a generalization of the traditional Jarzynski-W\'ojcik XFT. At the simplest level, it transitions to the traditional equality for zero initial correlations and energy conserving dynamics  - as it should. The breaking of the equality means the ratio of the forward and backward probabilities $R= \frac{\rm{Prob}[\Gamma(+q)]}{\rm{Prob}[\Gamma(-q)]}$ is now only located within a fixed, finite interval of size $\Delta R=e^{\Delta \beta q}(e^{-\Delta \II_u} - e^{ - \Delta \II_l})$, governed by the correlative structure in the initial quantum state. This is again to be expected, since in the absence of specifying finer details of the interaction dynamics we cannot a priori tell whether a particular interaction is sensitive to the correlations. Put another way, some interactions are better at activating the correlations than others, and as we increase the correlations we widen this finite interval. Equivalently, in the exponentiated XFT \textit{equality} in equation \eqref{eq_alt_NEeq}, this deviation is parameterised by the correlation factor $\eta$.

The increase of $\Delta R$ is exactly the distortion of the usual thermodynamic arrow, however it is important to note a distinction between the fluctuation theorem setting and the setting based on traditional expectation values. As already mentioned, when we measure heat-flow via $Q=\Tr [ H_A (U \rho U^\dagger - \rho)]$ we are not introducing any local measurement-disturbance into the system. Any entanglement present initially can influence the subsequent interactions and so can provide dramatic distortions of thermodynamic directionality. Indeed, for the most extreme case of a \emph{pure} multipartite state with local thermal states \emph{no} restriction exists beyond energy conservation and any such transformation can be done deterministically, including a maximal flow of heat from the colder to the hotter system (see \cite{partovi} and \cite{JR1} for details). 

Recall that any mixed state $\sigma_X$ admits a purification $\sigma_X \rightarrow \sigma_{XE} = |\psi\>_{XE}\< \psi|$, which is unique up to arbitrary unitaries on the purifying environment $E$. If one adopts this perspective, one has that for \emph{any} fixed thermal states $\rho_A$ and $\rho_B$, the issue of how large $\Delta R$ is amounts to asking how much of the purifying correlations is present in the state $\rho_{AB}$ for the composite system $AB$. Such states $\rho_{AB}$ range between the product state (molecular chaos) $\rho_A \otimes \rho_B$, and the situation where $\rho_{AB}= |\Psi_{AB}\>\< \Psi_{AB}|$, and $B$ is a purification of $A$.

\subsection{Going beyond sharp energy measurements}\label{beyond-sharp}
As mentioned, the sharp energy measurements $\M_1$ used are
quite destructive of coherence, and so one might wonder whether an XFT
can be obtained for more gentle POVMs. In other words, can we perform
the time-reverse pairing trick using mixed quantum states?

Given a preparation of some $\sigma_{AB}$ by the initial measurement
$\M_1$, we wish to do the pairing trick with the state $\sigma_{AB}$ and
a time-reversed twin. If we drop the assumption that $\M_2$ is a
sharp energy measurement, but leave it unspecified as $\M_2 =
\{M^{(2)}_{\phi'} \otimes N^{(2)}_{\chi'} \}$ we then require a
generalization of (\ref{dynamical}). Using the time-reversal invariance
of the unitary interaction we have $\Tr[M^{(2)}_{\phi'} \otimes
N^{(2)}_{\chi'} U \sigma_{AB} U^\dagger ] =\Tr[(\Theta \sigma_{AB}
\Theta^\dagger) U \Theta M^{(2)}_{\phi'} \otimes N^{(2)}_{\chi'}
\Theta^\dagger  U^\dagger ]$, and from this we see that, for the pairing
trick to work, the POVM elements of $\M_2$ must \emph{themselves be
valid quantum states} of the same form prepared by $\M_1$ and the set of
elements should be closed under the time-reversal operator
$\Theta$. (A similar requirement arises in the derivation of the non-equilibrium equality in section IV.) This on its own is a highly restrictive condition, and
explains why forming a theorem for more general POVMs than the
projective case is difficult.

\subsection{Application to a semi-classical maximum work theorem}\label{work-theorem}
The above results, and in particular (\ref{average}), find simple
application in a semi-classical maximum work theorem scenario
\cite{callen} in which a quantity of ordered energy is extracted from a
primary quantum subsystem $A$. The primary system is free to dump
entropy in the form of heat into a heat sink $B$, with fast relaxation
times, and exchange mechanical work with a third (classical) adiabatic
system $C$.

On the assumption of conservation of energy for the composite system
$ABC$ and the adiabaticity of $C$ the averaged relation (\ref{average})
leads to
\begin{eqnarray}
d W_C \le  - d U_A + T_B dS_A - T_B\< d\II\>
\end{eqnarray}
where $T_A dS_A := \<dq\>$ corresponds heat flowing into $A$, and we assume for simplicity that no net work is done on $A$. This does make the identification of $\Delta \epsilon$ with mechanical work, which can be debated as more or less sensible given that in extreme quantum regimes this can have broad distributions. We also make an identification of $S_A$ as the thermodynamic entropy, although again this is requires more care if the system is taken to finish out of equilibrium. We do not expand on these points here, but at the simplest level the main point of this application is to illustrate the contribution that the initial correlationsbetween the primary subsystem and the reversible heat sink provide to the usual maximum work theorem, and in the process illustrate the well-known
energetic value of correlations
\cite{szilard-1929,bennett82,feedback, sania}.

\subsection{Summary and outlook}
We have extended the Jarzynski-W\'{o}jcik Exchange Fluctuation Theorem
to the situation where we drop the assumption of molecular chaos, and
allow correlations to exist in the composite state. These correlations
results in a modification of the XFT relation and
can enhance the probability of heat flowing
in the backward direction. We have applied our results to deriving a
semi-classical maximum work theorem for correlated systems. Our work
highlights the difficulty of obtaining further results for situations
without initial and final measurements of energy. Our result show a deviation of the traditional XFT due to correlations present, and takes the form of a mutual information. A similar result has already been obtained for the case of the work Fluctuation Theorem \cite{feedbackcontrol} in which one allows feedback control. There the relevant mutual information is between the controller and the primary system. Furthermore the impact of this mutual information within the scenario has already been experimentally verified \cite{InfoHeatEngine}, and suggests that the generalized XFT obtained here should also be realisable in a similar manner with existing technologies.

\begin{acknowledgments}
D.J. is supported by the Royal Commission for the Exhibition of 1851. S.J. is supported by EPSRC grant
EP/K022512/1. T.R. is supported by the UK Engineering and Physical Sciences Research Council. Y. H. is supported by the Japan Society for the Promotion of Science for
 Young Scientists.  S. N. and M. M. are supported by Project for Developing Innovation Systems of
the Ministry of Education, Culture, Sports, Science and Technology (MEXT), Japan.

\end{acknowledgments}

\newpage
\appendix

\section{Derivation of equation (\ref{thermal})}
For any joint local POVMs $\{M_i\}$ on $A$ and $\{N_j\}$ on
$B$, we have defined the quantity  $\II(\rho_{AB}; M_i, N_j)$ via the expression
\begin{eqnarray}
\II(\rho_{AB}; M_i, N_j) &:=& \ln \left ( \frac{\Tr[ M_i \otimes N_j\rho_{AB}]}{(\Tr[M_i
\rho_{A}]\Tr[N_j \rho_{B}])} \right ) .
\end{eqnarray}

To show (\ref{thermal}) we consider the sharp energy measurement  $\M_1 = \{M_\phi
\otimes N_\chi \}$ where $\{ M_\phi=|\phi\>\<\phi| \}$ and
$\{\N_\chi=|\chi \>\<\chi| \}$ are the rank-1 projectors in the
local energy eigenbases.

For this particular initial measurement, the probability
$p(|\phi,\chi\>)$ of projecting into the state $|\phi, \chi\>$ under
$\M_1$ is simply given by $p(|\phi, \chi \rangle) = \< \phi, \chi | \rho_{AB} | \phi, \chi \rangle = \Tr [ M_\phi \otimes N_\chi \rho_{AB}]$. However, from the definition of $\II$ we have that
\begin{eqnarray*}
\Tr[M_\phi \otimes N_\chi \rho_{AB}] &=& e^{\II ( \rho_{AB} ; M_\phi, N_\chi)}\Tr[ M_\phi \rho_A] \Tr[N_\chi \rho_B].
\end{eqnarray*}
By assumption the state $\rho_{AB}$ has thermal marginals and so we have that
\begin{eqnarray*}
\Tr[ M_\phi \rho_A] \Tr[ N_\chi \rho_B] &=& e^{-\beta_A E_\phi -\beta_B E_\chi}/(Z_A Z_B).
\end{eqnarray*}
Substitution of these terms into $p(| \phi, \chi \rangle)$ gives
\begin{eqnarray*}
p(|\phi,\chi \>) &=& e^{-\beta_A E_\phi -\beta_B E_\chi - \log(Z_AZ_B) +
 \II[\rho_{AB}; M_\phi, N_\chi]},
\end{eqnarray*}
while the probability of obtaining $|\phi', \chi ' \>_*$ in the same measurement on $\rho_{AB}$ is given by 
\begin{eqnarray*}
p(|\phi',\chi' \>_*) &=& e^{-\beta_A E'_\phi -\beta_B E'_\chi - \log(Z_AZ_B) +
 \II[\rho_{AB}; M_{\phi'_*}, N_{\chi'_*}]}.
\end{eqnarray*}
Taking the ratio of these two probabilities leads us to the desired result
\begin{eqnarray}
p(|\phi,\chi\>) &=& p(|\phi',\chi'\>_*) e^{\Delta \beta q +\beta_B \Delta \epsilon - \Delta \II(\gamma)},
\end{eqnarray}
where $\Delta \beta = \beta_A -\beta_B$, and 
\begin{eqnarray*}
 \Delta \II(\gamma) &= &\II[\rho_{AB};M_{\phi_*'}, N_{\chi_*'} ] - \II[\rho_{AB};
M_\phi,N_\chi],
\end{eqnarray*} as claimed.

\section{Derivation of the abstract fluctuation theorem}

Here we fill in the details leading up to equation \eqref{eq_alt_NEeq}.

Using the discretised expression for $P(x)$ we have
\begin{eqnarray*}
P_X(x) e^{-x} &=& \sum_{\mu\nu} \delta(x-X_{\mu\nu}) p_{\mu\nu} e^{-x} \\
&=& \sum_{\mu\nu} \delta(x-X_{\mu\nu}) p_{\mu\nu} e^{-X_{\mu\nu}} \\
\label{f_munu}
& \equiv &  \sum_{\mu\nu} \delta(x-(-\wt{X}_{\mu\nu})) \wt{f}_{\mu\nu}\\
&:= &F_{\wt X}(-x).
\end{eqnarray*}
In the first to second we have used a property of delta functions, $g(x)\delta(x-x_0) = g(x_0)\delta(x-x_0)$, for some function $g(x)$, and in the second to third line we have defined $\wt{f}_{\mu\nu} := p_{\mu\nu} e^{-X_{\mu\nu}}$ and $\wt{X}_{\mu\nu} := -X_{\mu\nu}$. The third line is a probability density function for the new random variable $\wt{X}_{\mu\nu}$ if $\wt{f}_{\mu\nu} $ is a probability distribution.

We choose the random variable $X_{\mu\nu}$ to be given by
\begin{equation}
X_{\mu\nu} = \ln p_\mu - \ln f_\nu + \Delta \II_{\mu\nu}.
\end{equation}
where the correlation $$\Delta \II_{\mu\nu} = \II_\nu - \II_\mu=\ln \left(\frac{p_\mu}{p^0_\mu} \right)-\ln \left(\frac{f_\nu}{f^0_\nu}\right)$$ and $f_\nu=\Tr[M_\nu \otimes N_\nu \rho M^\dag_\nu \otimes N^\dag_\nu ]$, $f^0_\nu=\frac{e^{-\beta_A \phi'}}{Z_A}\frac{e^{-\beta_B \chi'}}{Z_B}$ are the probabilities of the final measurement on the correlated $\rho$ and product $\rho_A \otimes \rho_B$ states.
With this $X_{\mu\nu}$ we have simply $X_{\mu\nu}= \ln p^0_\mu  - \ln f^0_\nu$ and using the expressions for these uncorrelated probability distributions we obtain
\begin{equation}
X_{\mu\nu}
= \beta_A q^A_{\mu\nu} + \beta_B q^B_{\mu \nu}, 
\end{equation}
in terms of the sharp heat into $A$ and $B$ 
\begin{equation}
q^A_{\mu\nu} = \phi' - \phi,\quad q^B_{\mu\nu} = \chi' - \chi.
\end{equation}
%

Since $\delta(x-X_{\mu\nu}) e^{-X_{\mu\nu}} = \delta(x-X_{\mu\nu}) e^{-x}$, we are allowed to drop the $\mu,\nu$ labels convert $X_{\mu\nu}$ into the continuous random variable $x = \beta_A q^A + \beta_B q^B$. Define $q:= q^A$, $\Delta \epsilon := q^B+q$, and $\Delta \beta = \beta_A - \beta_B$, then 
with this we average the left hand side of the non-equilibrium equality
$\langle e^{-x} \rangle = \langle e^{-\Delta \beta q - \beta_B \Delta \epsilon} \rangle_p$, where the subscript $p$ indicates that the average is with respect to the $p_{\mu\nu}$ probability distribution given in equation \eqref{eq_pmunu}.

To calculate the correlation factor $\eta:= \langle F_{\wt X}(-x) \rangle$ 
let us formally write $\wt f_{\mu\nu} = f_\nu \wt f_{\mu|\nu}$. We have by definition $\wt f_{\mu\nu} = p_{\mu\nu} e^{-X_{\mu\nu}} $. 
Since $p_{\mu\nu} = p_\mu p_{\nu|\mu}$ and $e^{-X_{\mu\nu}} = \frac{f_\nu}{p_\mu}e^{-\Delta I_{\mu\nu}}$ we can deduce
\begin{equation}
 \wt f_{\mu|\nu} := p_{\nu|\mu}e^{-\Delta \II_{\mu\nu}}.
\end{equation}
Therefore
\begin{eqnarray}
F_{\wt X}(-x)  =  \sum_{\mu\nu} \delta(x+\wt{X}_{\mu\nu}) f_\nu p_{\nu|\mu}e^{-\Delta \II_{\mu\nu}}.
\label{Fv}
\end{eqnarray}

Is $f_\nu p_{\nu|\mu}$ a valid probability distribution? The $f_\nu$ part is fine since it is the probability of projecting the state $\rho$ onto $M_\nu\otimes N_\nu$.
The conditional $p_{\nu|\mu}$ is given in equation \eqref{p_cond}, note that it is equivalent to its time-reversed expression $|\langle \phi,\chi | U^\dag | \phi',\chi' \rangle |^2 =: f_{\mu|\nu} $, c.f. equation \eqref{dynamical}.
This is the probability that the state starts in $\ket {\phi'} \otimes \ket{\chi'} $, evolves under $U^\dag$ and is projected onto $\ket {\phi} \otimes \ket{\chi} $, and it satisfies $\sum_{\mu}f_{\mu|\nu} = 1$, therefore 
 $\sum_{\mu\nu} f_\nu f_{\mu|\nu} = 1$ and we do indeed have a valid probability $f_{\mu\nu} := f_\nu f_{\mu|\nu}$ (note carefully the difference between tildes and no tildes).

\section{Details for the toy example}

%

Consider first the generalised exchange fluctuation theorem in equation \eqref{genXFT}
\begin{align}
\hspace{-0.2cm} e^{\Delta \beta q +\beta_B\Delta \epsilon -\Delta \II_l} \le \frac{\mathrm{Prob}[\Gamma(q,\Delta \epsilon)]}{\mathrm{Prob}[\Gamma(-q, -\Delta \epsilon)]} \le e^{\Delta \beta q +\beta_B \Delta \epsilon-\Delta \II_u}.
\end{align}

We focus on the XFT for positive heat $q=1$ flows into $A$, in this example these are the histories $$\gamma[q=1|j]: \,\, (m,n)=(0,j) \rightarrow (m',n')=(1,j-1),$$ for $j=1,\dots,d-1$ and we have chosen the time-reversed state to be the spin-flipped one.
For these transitions, 
\begin{equation}
\label{eps0}
\Delta \epsilon = \<1,j-1|H_A+H_B|1,j-1\> -  \<0,j|H_A+H_B|0,j\> = 0.
\end{equation}
Note that the $\Delta \epsilon=0$ even for the reverse transition $\gamma[q=-1|j]$, and these are the only histories permitted by the interaction.

The correlation function for projective energy measurement $M_m \otimes N_n = \ket{m}\bra{m}\otimes \ket n\bra n$ is
\begin{align}
\mathcal I(\rho_{AB};m,n) &= \ln\left[\frac{\bra{mn}\rho_{AB}\ket{mn}}{\bra m \rho_A\ket m \bra n \rho_B\ket n}\right]\\
&= \beta_A m + \beta_B n + \ln \frac{\lambda_{mn}}{Z_A Z_B}
\end{align}
The sharp heat into $A$ is $q = \langle m'|H_A|m'\rangle - \langle m|H_A|m\rangle = m'-m = 0,\pm 1$ since $m = 0,1$.
The change in the correlation function is
\begin{align}
\Delta \II(\gamma[q=1|j]) &=\mathcal I(\rho_{AB};1,j-1) - \mathcal I(\rho_{AB};0,j)\\
&= \Delta \beta + \ln \frac{\lambda_{1,j-1}}{\lambda_{0,j}}
\end{align}
Later we will also make use of $$\Delta \II(\gamma[q=-1|j]) = -\Delta \beta + \ln \frac{\lambda_{0,j}}{\lambda_{1,j-1}} = -\Delta \II(\gamma[q=1|j]).$$

The upper $u$ and lower $l$ bounds on $\Delta \mathcal I$ are given by $\Delta \mathcal I_u = \Delta \beta + \max_j \left\{\ln\frac{\lambda_{1,j-1}}{\lambda_{0,j}}\right\}_{j=1}^{d-1}$ and $\Delta \mathcal I_l = \Delta \beta + \min_j\left \{\ln\frac{\lambda_{1,j-1}}{\lambda_{0,j}}\right\}_{j=1}^{d-1}$.

Substituting these expressions into equation \eqref{genXFT} we obtain
\begin{equation}
\label{eg_XFT}
\min\left\{\frac{\lambda_{0,j}}{\lambda_{1,j-1}}\right\} \leq \frac{\mathrm{Prob}[\Gamma(q=1)]}{\mathrm{Prob}[\Gamma(q=-1)]} \leq \max\left\{\frac{\lambda_{0,j}}{\lambda_{1,j-1}}\right\} 
\end{equation}

The second thermodynamic inequality is simply $\Delta \beta \< q \> \geq \langle \Delta \mathcal{I} \rangle$ because $\Delta \epsilon = 0$ for this energy-conserving interaction. The average difference of the correlation function is
\begin{align}
\langle\Delta \mathcal I \rangle &= \sum_j (\mathrm{Prob}[q=\pm1|j] \Delta \mathcal I(\gamma[q=\pm1|j])\\
&= \sum_j (\lambda_{0j}-\lambda_{1,j-1})( \Delta \beta + \ln \frac{\lambda_{1,j-1}}{\lambda_{0,j}})\sin^2\omega_jt 
\end{align}



Let us now turn our attention to the correlation factor 
$$\eta = \langle e^{-\Delta \II} \rangle_f$$
from equation \eqref{eq_alt_NEeq}. 
The initial and final measurement labels are  $\mu = (0,j)$ and $\nu = (1,j-1)$, we have $\Delta \II_{\mu\nu} = \Delta \beta + \ln\frac{\lambda_{1,j-1}}{\lambda_{0j}}$, and the probability $f_{\mu\nu} = f_\nu f_{\mu|\nu} = \sum_j\lambda_{1,j-1}\sin^2\omega_j t$ since $ f_{\mu|\nu} = |\langle 0,j| U^\dag | 1,j-1 \rangle |^2 $ and $f_\nu = \bra{1,j-1}\rho_{AB}\ket{1,j-1}$. Including also the remaining transitions $\mu = (1,j-1)$ and $\nu = (0,j)$, we obtain
\begin{align}\nonumber
\langle e^{-\Delta \II} \rangle_f &= \sum_{j=1}^{d-1}(\lambda_{1,j-1}e^{-\Delta \beta - \ln\frac{\lambda_{1,j-1}}{\lambda_{0j}}}\\ 
&\quad\quad+\lambda_{0j}e^{\Delta \beta - \ln\frac{\lambda_{0j}}{\lambda_{1,j-1}}})\sin^2\omega_j t
\end{align}
and this may be simplified to give the $\eta$ in the main text.

\end{document}